\documentclass[12pt,fleqn]{article}
\usepackage{xiiiemc}
\usepackage{natbib}
\usepackage{fancyhdr}

\begin{document}

\title{QUANTUM NOISE AND ITS IMPORTANCE TO THE QUANTUM CLASSICAL TRANSITION PROBLEM: ENERGY MEASUREMENT ASPECTS}
\author{\textrm{%
\begin{tabular}{l}
\textbf{Ad\'elcio C. Oliveira}- e-mail adelcio@ufsj.edu.br \\
Departamento de F\'isica e Matem\'atica, Universidade Federal de
S\~ao
Jo\~ao Del Rei, \\
Ouro Branco, 36420 000, Minas Gerais , Brazil \\
\textbf{Zolacir T. Oliveira Junior, Nestor Santos Correia }
\\
Departamento de Ci\^encias Exatas e Tecnol\'ogicas, Universidade
Estadual de
Santa Cruz, \\
Ilh\'eus, 45662 000, Bahia, Brazil \\
\end{tabular}%
}}

\maketitle
\thispagestyle{fancy}

\lhead{\textcolor{blue}{\fontsize{9}{0pt}\usefont{T1}{phv}{m}{n}\it
XVI Encontro de Modelagem Computacional \\ \vspace{1.0pt} IV
Encontro Ci\^encia e Tecnologia de Materiais \\ \vspace{1.0pt} III
Encontro Regional de Matem\'atica Aplicada e Computacional \\
\vspace{-4.0pt} Universidade Estadual de Santa Cruz (UESC),
Ilh\'eus/BA, Brasil. 23-25 out. 2013.}} \chead{} \rhead{} %
\renewcommand{\headrulewidth}{0.0pt} \fancyfoot{}

\begin{abstract}
In the present contribution we discuss the role of experimental limitations
in the classical limit problem. We studied some simple models and found that
Quantum Mechanics does not re-produce classical mechanical predictions,
unless we consider the experimental limitations ruled by uncertainty
principle. We have shown that the discrete nature of energy levels of
integrable systems can be accessed by classical measurements. We have
defined a precise limit for this procedure. It may be used as a tool to
define the classical limit as far as the discrete spectra of integrable
systems are concerned.
\end{abstract}


\thispagestyle{fancy}




\section{Introduction}

 There were no conclusive explanation for the disappearance of
quantum effects in the macroscopic regime. Which of the quantum
features we don't see in our day life experience? Does quantum
mechanics reproduce the observed results of macroscopic experiments?
The first try to answer this question is attributed to Bohr, the
correspondence principle. Bohr's correspondence principle
\citep{Bohr}. In this contribution, the main goal is to investigate
the role of quantum noise in the quantum-classical transition
problem. We focuss on the spectra discreteness. We propose a
procedure to use spectroscopic information from model Hamiltonians
and time energy Heisenberg relations in order to decide whether a
quantum system can be described by CM. The quantum behavior is
characterized by the discreteness of energy spectra. We are
considering a gedankenexperiment, where the experimentalist does not
know Quantum Mechanics but tries obtain the energy spectrum. Our
results confirm previous one \citep{renato2007,adelcio2012} that
asserts the importance of the quantum noise to understand quantum
classical transition problem.
\newpage
\section{\protect\bigskip The quantum noise in position and momentum
measurements}

In this section we show that quantum uncertainty are inherent to any system.
After a position measurement we assume that the quantum state is $\left\vert
x\right\rangle .$ We prepare an ensemble of N identical particles prepared
in the same initial state, then the result of the measurement of position
for i-th particle is $x_{i}$ \ and its mean is $\overline{x_{i}}=r$, thus $%
x_{i}=r+\xi _{i},$ where $\xi _{i}$ is $N(0,\delta x^{2}),$ i.e. normal
random variable with variance $s^{2},$ thus the state after the measurement
is%
\begin{equation}
\left\vert \varphi \right\rangle =\overline{\left\vert x\right\rangle },
\end{equation}%
the overbar represents ensemble mean. Using the position displacement
operator, $\widehat{S}(\lambda )=\exp \left( -\frac{i}{\hbar }\widehat{p}%
\lambda \right) $ , and the resolution of unity in $\left\vert
p\right\rangle $ bases we have\bigskip
\begin{equation}
\left\vert \varphi \right\rangle =\int dp\overline{\exp \left( -\frac{i}{%
\hbar }p\xi _{i}\right) }\left\vert p\right\rangle \left\langle p\right.
\left\vert r\right\rangle .
\end{equation}%
This ensemble mean can be easily computed \citep%
{Oksendal,Gardiner,Oliveira2013}, it is $\overline{\exp \left( -\frac{i}{%
\hbar }p\xi _{i}\right) }=\exp \left( -\frac{p^{2}\Delta x^{2}}{2\hbar ^{2}}%
\right) ,$ also we have $\left\langle p\right. \left\vert r\right\rangle
=(2\pi \hbar )^{-1/2}\exp \left( -\frac{ipr}{\hbar }\right) ,$ them%
\begin{equation}
\left\vert \varphi \right\rangle =(2\pi \hbar )^{-1/2}\int dp\exp \left( -%
\frac{p^{2}\delta x^{2}}{2\hbar ^{2}}\right) \exp \left( -\frac{ipr}{\hbar }%
\right) \left\vert p\right\rangle .
\end{equation}%
and
\begin{equation}
\varphi (x)=\left\langle x\right. \left\vert \varphi \right\rangle =(2\pi
\hbar )^{-1}\int dp\exp \left( -\frac{p^{2}\delta x^{2}}{2\hbar ^{2}}\right)
\exp \left( -\frac{ip(r-x)}{\hbar }\right)  \label{gaus1}
\end{equation}%
integrating (\ref{gaus1}) we obtain
\begin{equation}
\varphi (x)=\frac{1}{\sqrt{2\pi s^{2}}}\exp \left( -\frac{(x-r)^{2}}{2\delta
x^{2}}\right) .  \label{gaus2}
\end{equation}%
For the momentum measurement we observe that $\widehat{T}(\lambda )=\exp
\left( \frac{i}{\hbar }\widehat{q}\lambda \right) $ is the momentum
displacement operator and $p_{i}=d+\zeta _{i},$ $\ $ again $\zeta _{i}$ is $%
N(0,\delta p^{2})$ then
\begin{equation}
\varphi (p)=\frac{1}{\sqrt{2\pi a^{2}}}\exp \left( -\frac{(p-d)^{2}}{2\delta
p^{2}}\right) ,  \label{gaus3}
\end{equation}%
where $d$ is the ensemble mean of $p_{i}.$The maximum precision is
achieved when $\delta x\delta p=\hbar /2,$ this is the standard
quantum limit (SQL) \citep{Lynch,Appleby}. Even if one measures the
state with greater precision than SQL \citep{Rozena}, one can not
use it as an initial state, as pointed by Ballentine
\citep{Ballentine 1970} \textquotedblleft the uncertain principle
restricts the degree of statistical homogeneity which is possible to
achieve in an ensemble of similarly prepared systems and thus it
limits the precision for any system can made".

\newpage
\section{\protect The large quantum numbers limit}

It is well known that time and frequency are conjugated variables in
a pair of Fourier transforms in classical physics, the duration of a
signal and the respective frequency are subjected to an unsharpness
relation ($\Delta t.\Delta \omega \geq \frac{1}{2}$) that is
classical indeed. Frequency, in quantum theory, is another way of
speaking about energy. An uncertainty relation between time and
energy must be seriously considered, studied and interpreted,
although there are objections due to the fact that time is not
associated to a dynamical operator canonically conjugated to the
hamiltonian. An exhaustive examination of this matter is made by
Peres in his book \citep{Peres 2002}. On the other hand we have to
assume that QM is not an objective description of physical reality.
It only predicts the probability of occurrence of \textit{stochastic
macroscopic events}, following specified preparation procedures.

The main stream of our proposal is not face the discussion of the
time-energy uncertainty relation, we face the problem of to decide
under what conditions a system may be considered as Classical or
Quantum. It is in this sense that we tackle the problem of dealing
with the physical reality: how can one measure and what is in fact
measured. Thus, we use the product of energy differences between
neighbor levels by the corresponding classical period differences to
compare with the time-energy uncertainty relation to classify a
system as classical or quantum. If this product fulfills the
time-energy uncertainty relation it is quantum otherwise it is
classical.

To do this we add an element connected to this subject that concerns
integrable systems: the decision wether a system is classical or quantum
depends on the experimental apparatus which are essentially classical. One
undoubtedly quantum feature is the discreteness of at least part of the
energy spectrum. The essential idea here is that if one tries to measure the
energy of the system in question using classical canonical pairs $q$, $p$,
the information about the quantum nature of the particle will be lost. In
spite of this fact, as we show in what follows, the function
\begin{equation}
y(n)=\left\vert \Delta E_{n}\Delta \tau _{n}\right\vert  \label{yfunction}
\end{equation}%
where $\Delta E_{n}=\left( E_{n}-E_{n-1}\right) /2$ is merely the
energy difference between two neighbor levels, (it is the maximum
uncertainty in energy for the state $a\left\vert E_{n}\right\rangle
+b\left\vert E_{n-1}\right\rangle $ ) and $\Delta \tau _{n}=\left(
\tau _{n}-\tau _{n-1}\right) /2,$ with $\tau _{n}$ being the
classical period associated to the energy $E_{n}.$ If the experiment
has an accuracy $\delta t$ it limits
the period measurement precision, in real systems it is desirable that $%
\delta t<<\Delta \tau $, we assume that they are of same order, $\Delta \tau
\approx $ $\delta t.$

The $y(n)$ function can be heuristically justified if we consider the
Bohr-Sommerfeld quantization rule for periodic systems, that states
\begin{equation}
I=\oint pdq=2\pi \hbar n
\end{equation}%
thus we have%
\begin{equation}
\left\langle K\right\rangle =\frac{\pi n\hbar }{\tau },  \label{ktau}
\end{equation}%
where $\left\langle K\right\rangle $ is the the mean kinetic energy, $\tau $
being the classical period associated with $K$ and $n$ is a quantum number.
\newpage
Observing that
\begin{equation}
\delta I=\oint \delta Hdt
\end{equation}%
Consider two neighboring periodic motions of the same system. Then H
is a constant, and we have
\begin{equation}
\delta I=\tau \delta H.  \label{deltaH}
\end{equation}%
Note that (\ref{deltaH}) determines that the energy of the system

depends only on I, then Bohr's quantiztion rule determines the
energy of the system.
The Bohr-Sommerfeld quantization rule makes a
direct conection between
kinetics energy and classical period, it was also demonstrated \citep%
{adelcioJMP} that, for integrable systems, is possible to
reconstruct semiclassicaly the quantum state using classical
dynamics. Since you know the Wigner function of the system it is
possible to infer the period related to the state \citep{adelcioJMP,
Oliveira01}. \bigskip Alternatively, we can use the semiclassical
quantization rule \citep{Berryi, Stockmann} for integrable systems,
we have a similar result \citep{Stockmann, Wisniacki, Novaes}. This
function (\ref{yfunction}) is stated for a more clear definition of
a classical limit. As we will show, the fact that one will only be
able to see a continuum of energies does not mean that the classical
limit has been reached. It only reveals that we have become
\textquotedblleft myope\textquotedblright\ to see the nature of the
spectrum.

In what follows, we will consider a simple one dimensional systems with
discrete spectra. Let us assume that a state has been prepared in the
following way
\begin{equation}
\left\vert \Psi (0)\right\rangle =a\left\vert E_{n}\right\rangle
+b\left\vert E_{n-1}\right\rangle  \label{1}
\end{equation}%
where $\left\{ \left\vert E_{n}\right\rangle \right\} $ are eigenstates of
the hamiltonian $H$
\begin{equation}
\hat{H}\left\vert E_{n}\right\rangle =E_{n}\left\vert E_{n}\right\rangle
\label{2}
\end{equation}%
The variance in energy difference, for any time, of this state is given by
\begin{equation}
\Delta E=\left\vert a\right\vert \left\vert b\right\vert \left[ E_{n}-E_{n-1}%
\right]
\end{equation}%
Using the normalization condition $\left\vert a\right\vert ^{2}+\left\vert
b\right\vert ^{2}=1$, it is easy to check that the maximum for this energy
difference is achieved for $\left\vert a\right\vert =\left\vert b\right\vert
=1/\sqrt{2}$, and is given by
\begin{equation}
\Delta E(n)=1/2\left\vert E_{n}-E_{n-1}\right\vert .
\end{equation}
How do we define the classical limit in this simple case? Let us
assume the following $\displaystyle\lim_{n\rightarrow \infty
}\frac{\Delta E_n}{E_n}=0$. The above expression means that as the
energy increases $(E_{n})$, the uncertainty in energy of the state
$\left| \Psi (t)\right\rangle $ becomes negligible as compared to
$E_{n}.$ Let us discuss a few examples.
\newpage

\emph{Harmonic Oscillator}

For the harmonic oscillator we have
\begin{equation}
\Delta E=\frac{1}{2}\hbar \omega \Longrightarrow \frac{\Delta E}{E_{n}}=%
\frac{1/2}{(n+1/2)}  \label{delta e oscilador harmonico}
\end{equation}

and we clearly have $\displaystyle\lim_{n\rightarrow \infty }\frac{\Delta E_n%
}{E_n}=0$.

\emph{Particle in a box}

In this case we have
\begin{equation}
E_{n}=\hbar ^{2}\frac{n^{2}\pi ^{2}}{2ma^{2}},
\end{equation}%
where $n=1,2,3,...$ and $m$ is the mass of the particle and $a$ is
the box width. Then we have
\begin{equation}
\Delta E_{n}=\hbar ^{2}/4\left\vert \frac{(2n-1)\pi ^{2}}{ma^{2}}\right\vert
.  \label{Delta E caixa}
\end{equation}%
Once again we have $\lim_{n\rightarrow \infty
}{\displaystyle{\frac{\Delta E_{n}}{E_{n}}}}\rightarrow 0.$ The fact
that this limit is zero is the argument usually found in textbooks
in order to define the classical limit\citep{Home,Cohen,Eisberg}. Of
course it is true that when one considers the high energy limit, it
becomes increasingly difficult to obtain good experimental
resolution. However, as we show below, this limit does not
necessarily imply that it is impossible to obtain the needed
resolution. As we know, quantization has been observed
in some macroscopic systems like superconducting Josephson junctions \citep%
{Richard}. Besides, one can use something analogous to the
Heisenberg microscope \citep{Quantummesurement} in order to obtain
the velocity of the particle. In the case of a particle in a box,
discussed above, once we know the velocity, the energy is easily
obtained. The measurement of the velocity implies a perturbation of
the position which depends on the characteristics of the apparatus
and in principle, is independent of the quantum number $n$. Thus, if
one uses the adequate measurement the discrete character of a
spectrum can be verified.

\section{Classical measurement of the Energy}

When one deals with realistic systems, say, atoms, the spectrum is obtained
from the emitted or absorbed electromagnetic radiation. A completely
different approach is used for macroscopic systems. Within CM, for closed
systems, the energy is a function of position and momentum $E=f(p,q).$ So,
from now on, we will call Classical Measurement of energy every process
which makes use of the relation $E=f(p,q)$ in order to obtain the energy of
a given system, classical or quantum. As an example, we may look at the
particle in a box again. In this case, the energy is a direct function of
the velocity, so that once we determine de velocity, the energy will be
defined. In practice, one may measure the time it takes for $2s$ inversions
in the momentum and so determine the period of the motion .
\newpage
 Classically, the period is given by
\begin{equation}
\tau =\frac{2a}{v}=a\sqrt{\frac{2m}{E}.}
\end{equation}%
In the above expression, we are considering the correspondent classical
period for a specific energy eigenvalue. Since the energy levels are
determined by QM, regardless of the energy scale one is talking about, the
allowed values of the period are
\begin{equation}
\tau _{n}=\frac{2a^{2}m}{\hbar n\pi }.
\end{equation}%
So, the difference in period for two quantum neighboring levels is
\begin{equation}
\Delta \tau =\frac{\tau _{n}-\tau _{n-1}}{2}=-\frac{a^{2}m}{\hbar \pi }\frac{%
1}{\left( n-1\right) n}.  \label{delta tau}
\end{equation}

From the above expression one can see that it becomes increasingly difficult
to distinguish two higher adjacent energy levels by this method. Let us take
a look at the product $\left\vert \Delta E_{n}\Delta \tau _{n}\right\vert $
where $\Delta E_{n}=\left( E_{n}-E_{n-1}\right) /2$ and $\Delta \tau
_{n}=\left( \tau _{n}-\tau _{n-1}\right) /2.$ $\tau _{n}$ is the classical
period associated with the energy $E_{n}.$ It is interesting to observe the
behavior of the function
\begin{equation}
y(n)=\left\vert \Delta E_{n}\Delta \tau _{n}\right\vert .
\end{equation}%
There is no mathematical restriction for $y(n)\,,$ but the set
$\{n\in |y(n)<\hbar /2\}$ is beyond the SQL since Quantum Mechanics
forbids such precision\footnote[5]{\textsc{The time-energy
uncertainty relation} $\Delta E\Delta t\geq \hbar /2$ is a first
principle limitation and has nothing to do with experimental errors.
A deeper discussion on the subject can be found in Refs.
\citep{Quantummesurement,Appleby,Rajeev}}. For the case in question,
\begin{equation}
y(n)=\frac{\hbar \pi }{4}\frac{(2n-1)}{\left( n-1\right) n}.
\end{equation}%
If we take $n\geq 4,$ we find $y(4)<\hbar /2$ . Here, we are just using the
fact that the time-energy uncertainty relation has to be respected since we
are considering that Quantum Mechanics must prevent Classical Mechanics.
Also it is easy to see that $\displaystyle\lim_{n\rightarrow \infty }y(n)=0.$

In the more realistic case of a hydrogenoid atom we have
\begin{equation}
E_{n}=-\frac{\mu Z^{2}e^{4}}{2\hbar ^{2}n^{2}}  \label{E 1/r}
\end{equation}%
where $Ze$ is the total charge interacting with the electron of charge $e$,
and $\mu $ is the reduced mass of the system. As opposed to the previous
case, the energy levels become closer as $n$ grows. Thus, the question is:
\textit{from which $n$ can we say that the spectrum is continuous }from the
point of view of Classical Mechanics?

\newpage

In this case
\begin{equation}
\Delta E=\frac{\mu Z^{2}e^{4}}{4\hbar ^{2}}\left\vert \frac{2n-1}{%
n^{2}(n-1)^{2}}\right\vert ,  \label{Delta E 1/r}
\end{equation}%
\begin{equation}
\Delta T=\frac{\pi \hbar ^{3}}{Z^{2}e^{4}\mu }\left[ 3n^{2}-3n+1\right] ,
\label{deta T 1/r}
\end{equation}%
and
\begin{equation}
y(n)=\frac{\pi \hbar \left( 2n-1\right) \left[ 3n^{2}-3n+1\right] }{%
4n^{2}(n-1)^{2}}.
\end{equation}%
Now, for the $1/r$ potential, we have $y(n)<\hbar /2$ for $n\geq 9$, so one
won't be able to observe the quantization for $n$ bigger than 9 while using
classical measurement of the energy.

Other interesting case is the Morse Potential, frequently used to
describe the spectra of molecules \citep{Oliveira01} . The Morse
potential is defined as
\begin{equation}
U(x)=D\left( e^{-2\alpha x}-2e^{-\alpha x}\right) ,
\end{equation}%
where $D$ and $\alpha $ are constants experimentally determined. For $s$
waves, i.e., the orbital angular momentum is zero, we have

\begin{equation}
E(n)=-D+\hbar \omega \left[ \left( n+1/2\right) -\frac{1}{\zeta }\left(
n+1/2\right) ^{2}\right] .
\end{equation}%
$\zeta $ is also experimentally determined. The classical period
corresponding to each $E(n)$ is given by
\begin{equation}
\tau (n)=2\pi \sqrt{\frac{MRo^{2}}{2\left\vert E(n)\right\vert \alpha ^{2}}}
\end{equation}%
where $Ro$ is a function of $\alpha $ and $\zeta .$ It is easy to verify
that, for the hydrogen molecule parameters, $y(n)<\hbar /2$ for all possible
$n$. Since the Morse potential is quasi-harmonic for low energies we observe
that it is not possible to distinguish neighboring discrete states with a
classical measurement. From this example, we may conclude that any potential
that is approximately harmonic have no assessable discrete spectrum through
a classical measurement of the energy.

\subsection{Harmonic Oscillator}

For a harmonic oscillator, $E=\frac{p^{2}}{2m}+\frac{1}{2}kq^{2}$\ , its
period is $\tau =\frac{2\pi }{\omega }$ where $\omega =\sqrt{\frac{k}{m}}$
which is energy independent, thus we conclude that it can not be used to
characterize the spectrum. Another way of classically determining the energy
can be obtained by measuring q and p at same time, therefore classical
energy uncertain is given by
\begin{equation}
\delta E=\frac{p}{m}\delta p+kq\delta q
\end{equation}

\newpage

Without lost of generality, we choose $\delta p=\sqrt{\frac{m\hbar \omega }{2%
}}a$ and $\delta q=\sqrt{\frac{\hbar }{2m\omega }}a$ then we obtain the
relation
\begin{equation}
\delta p\delta q=a^{2}\hbar /2.
\end{equation}

In case of a=1 we have the minimum uncertain defined under Hisenberg
relation, in general we have
\begin{equation}
\delta E=\left[ |p|\sqrt{\frac{\hbar \omega }{2m}}+k|q|\sqrt{\frac{\hbar }{%
2m\omega }}\right] a
\end{equation}%
then
\begin{equation}
\delta E^{2}=\frac{2}{\hbar }\left[ p\sqrt{\frac{\hbar \omega }{2m}}+kq\sqrt{%
\frac{\hbar }{2m\omega }}\right] ^{2}\delta p\delta q
\end{equation}%
From (\ref{delta e oscilador harmonico}) we have

\begin{equation}
\delta E=\frac{1}{2}\hbar \omega
\end{equation}%
taking into account that its energy is $E=\hbar \omega (n+1/2)$, then after
some straightforward algebra we find that $\delta p\delta q<\hbar /2$ for
all $n$, then we need an experimental resolution that is \ beyond SQL. Thus
we can say that his spectrum can not be resolved by Classical Energy
Measurement.

\section{Conclusion}

In this work we have shown that the discrete nature of the energy levels can
be accessed by classical measurements in some cases. We also defined a
precise limit for this procedure using the function $y(n)=\left\vert \Delta
E_{n}\Delta \tau _{n}\right\vert $ and comparing it with the time-energy
uncertainty principle. This maneuver gives us a complementarity principle
and a well defined mathematical limit dictated by the experiment. Of course,
the fact that we are not able to recognize the discrete nature of a spectrum
does not necessarily mean it is not discrete. It only means how
\textquotedblleft myope\textquotedblright\ we are, suggesting that Classical
Mechanics can be viewed as a blurring of essential aspects of Quantum
Mechanics and also explains why it took so long to find quantum effects.


\begin{thebibliography}{0}
\expandafter\ifx\csname natexlab\endcsname\relax\def\natexlab#1{#1}\fi
\expandafter\ifx\csname bibnamefont\endcsname\relax
  \def\bibnamefont#1{#1}\fi
\expandafter\ifx\csname bibfnamefont\endcsname\relax
  \def\bibfnamefont#1{#1}\fi
\expandafter\ifx\csname citenamefont\endcsname\relax
  \def\citenamefont#1{#1}\fi
\expandafter\ifx\csname url\endcsname\relax
  \def\url#1{\texttt{#1}}\fi
\expandafter\ifx\csname urlprefix\endcsname\relax\def\urlprefix{URL }\fi
\providecommand{\bibinfo}[2]{#2}
\providecommand{\eprint}[2][]{\url{#2}}

\end{thebibliography}


\newpage

\begin{thebibliography}{99}



\bibitem[Angelo, 2007]{renato2007}
 Angelo, R. M., 2007.
\newblock ,Phys.Rev. A., \textbf{76}, 052111.

\bibitem[Appleby, 1998]{Appleby}
  Appleby, D. M., 1998.
  \newblock International Journal of Theoretical
 Physics,\textbf{37}, 1491.

\bibitem[Ballentine, 1970]{Ballentine 1970}
Ballentine, L. E., 1970.
\newblock Rev. Mod. Phys., \textbf{42}, 358.

\bibitem[Berry \& Balazs, 1978]{Berryi}
 Berry, M. V., Balazs, N. L., 1978.
\newblock J. Phys. A, \textbf{12} 625.

\bibitem[Bohr, 1998]{Bohr}
Bohr, N. , 1998.
\newblock \textit{Causality and Complementarity, supplementary
papers edited by Jan Faye and Henry Folse as The Philosophical
Writings of Niels Bohr},
\newblock Vol. IV, Woodbridge: Ox Bow Press.

\bibitem[Braginsky \&Khalili, 1992]{Quantummesurement}

Braginsky, V. B., Khalili, F. Y., 1992.
\newblock \textit{Quantum Measurement}
\newblock (Cambridge University Press, Cambridge).

\bibitem[Cohen et~al., 1977]{Cohen}
Cohen-Tannouji,C.,  Diu, B., Lalo{\"e},F., 1977.
\newblock \textit{\ Quantum Mechanics},
\newblock vol. 1 (Wiley, New York).

\bibitem[Eisberg \& Resnick, 1994]{Eisberg}
 Eisberg, R. M., Resnick,R., 1994.
 \newblock \textit{F\'{\i}sica Qu\^antica: \'Atomos, Mol\'eculas, S\'olidos, N\'ucleos e Part\'{\i}culas}
\newblock(Campus, S{\~{a}o} Paulo).

\bibitem[Gardiner \& Zoller, 2010 ]{Gardiner}
 Gardiner, C. W., Zoller, P. 2010.
 \newblock\textit{Quantum Noise}
 \newblock (Springer-Verlag, Berlin,Heidelberg).

\bibitem[Home, 1997]{Home}
 Home,D., 1997.
  \newblock\textit{Conceptual Foundations of Quantum Physics. An Overview from Modern Perspectives}
  \newblock (Plenum Press, New York and London).

\bibitem[Lynch, 1985]{Lynch}
 Lynch,R., 1985.
 \newblock Phys. Rev. Lett., \textbf{54}, 1599.

\bibitem[Novaes \& Aguiar, 2005]{Novaes}
 Novaes,M., Aguiar, M. A. M., 2005.
 \newblock Phys. Rev. A, \textbf{71} 012104.

\bibitem[Oksendal, 2000]{Oksendal}
 Oksendal, B., 2000.
 \newblock \textit{Stochastic Differential Equations: An introduction with applications, }
 \newblock 5 ed.,Springer-Verlag Heidelberg New York.

\bibitem[Oliveira \& Nemes, 2001]{Oliveira01}
 Oliveira, A.~C., Nemes, M.~C., 2001.
 \newblock Physica Scripta, \textbf{64}, 279.

\bibitem[Oliveira et~al., 2012]{adelcio2012}
 Oliveira,A. C., Bosco de Magalh\~{a}es, A.R., Peixoto Faria,J. G.,
 2012.
 \newblock Physica A, \textbf{391, }5082.

\bibitem[Oliveira, 2012]{adelcioJMP}
Oliveira, A. C., 2012.
\newblock Jour. Mod. Phys., \textbf{3, }694.

\bibitem[Oliveira, 2013]{Oliveira2013}
Oliveira, A. C., 2013.
 \textit{Classical Limit of Quantum Mechanics Induced by Continuous Measurements, }%
\newblock Submited Physica A.

\bibitem[Peres, 2002]{Peres 2002}
Peres, A., 2002.
\newblock \textit{Quantum
Theory: Concepts and Methods }
\newblock(Kluver Academic Publishers, New York, Boston, Dordrecht, London, Moscow), chapter 12, item 12-8 and
references therein cited.

\newpage

\bibitem[Rajeev, 2003]{Rajeev}
 Rajeev, S.~G., 2003.
 \newblock \textit{ A theory of
errors in quantum measurement},
\newblock arXiv: quant-ph/0306037.

\bibitem[Richard et~al., 1981]{Richard}
Richard, F. V., Web,R.A. 1981.
\newblock Phys. Rev. Lett., \textbf{47},265.

\bibitem[Rozena et~al., 2012]{Rozena}
 Rozema, L. A., Darabi, A., Mahler., D. H., Hayat, A., Soudagar, Y.,
Steinberg, A. M., 2012.
\newblock Phys. Rev. Lett., \textbf{109}, 100404.


\bibitem[Stockmann, 2000]{Stockmann}
St\"{o}ckmann, H. J., 2000.
\newblock \textit{Quantum Chaos: an introduction,}
\newblock Cambridge University Press, New York.

\bibitem[Wisniacki et~al., 2006]{Wisniacki}
 Wisniacki, D. A., Vergini, E., Benito,R. M., Borondo,F., 2006.
\newblock Phys. Rev. Lett., \textbf{97} 094101.



\end{thebibliography}
\end{document}